\RequirePackage{lineno}
\documentclass[aps,prl,twocolumn,showpacs,preprintnumbers,amsmath,amssymb,superscriptaddress]{revtex4}
\usepackage{graphicx}
\usepackage{dcolumn}
\usepackage{bm}
\usepackage{lineno}
\usepackage{color}
\begin{document}
\begin{setpagewiselinenumbers}
\begin{linenumbers}
\title{Evolution of the differential transverse momentum correlation function with centrality in Au+Au collisions at $\sqrt{s_{NN}} = 200$ \mbox{GeV}}
\collaboration{STAR Collaboration}\noaffiliation
\author{H.~Agakishiev}\affiliation{Joint Institute for Nuclear Research, Dubna, 141 980, Russia}
\author{M.~M.~Aggarwal}\affiliation{Panjab University, Chandigarh 160014, India}
\author{Z.~Ahammed}\affiliation{Lawrence Berkeley National Laboratory, Berkeley, California 94720, USA}
\author{A.~V.~Alakhverdyants}\affiliation{Joint Institute for Nuclear Research, Dubna, 141 980, Russia}
\author{I.~Alekseev~~}\affiliation{Alikhanov Institute for Theoretical and Experimental Physics, Moscow, Russia}
\author{J.~Alford}\affiliation{Kent State University, Kent, Ohio 44242, USA}
\author{B.~D.~Anderson}\affiliation{Kent State University, Kent, Ohio 44242, USA}
\author{C.~D.~Anson}\affiliation{Ohio State University, Columbus, Ohio 43210, USA}
\author{D.~Arkhipkin}\affiliation{Brookhaven National Laboratory, Upton, New York 11973, USA}
\author{G.~S.~Averichev}\affiliation{Joint Institute for Nuclear Research, Dubna, 141 980, Russia}
% ...
Version 3.3, Aug 12th, 2011, 11:59 AM
\begin{linenomath*}
\begin{abstract}
We present first measurements of the evolution of the differential transverse momentum correlation function, {\it C}, with collision centrality in Au+Au interactions at $\sqrt{s_{NN}}~=~200$ \mbox{GeV}.  This observable exhibits a strong dependence on collision centrality that is qualitatively similar to that of number correlations previously reported. We use the observed longitudinal broadening of the near-side peak of {\it C} with increasing centrality to estimate the ratio of the shear viscosity to entropy density, $\eta/s$, of the matter formed in central Au+Au interactions. We obtain an upper limit estimate of $\eta/s$ that  suggests that the produced medium has a small viscosity per unit entropy.   
\end{abstract}
\end{linenomath*}
\maketitle
\setcounter{page}{1}
Measurements carried out at the Relativistic Heavy Ion Collider (RHIC) during the last decade indicate that a strongly interacting quark gluon plasma (sQGP) is produced in heavy nuclei collisions at very high beam energies \cite{Ref1}. 
It has emerged that this matter behaves as a ``nearly perfect liquid", i.e., a fluid which has a very small shear viscosity per unit of entropy \cite{Ref1,Ref2}. It is a fascinating observation that the medium produced in relativistic heavy ion collisions reaches exceedingly large temperatures, of the order of $2 \times 10^{12}$ $\mbox{K}$ ~\cite{QGPtemperature}, in stark contrast to the very low temperature, $T < 3$ $\mathrm{K}$, required to achieve superfluid ${}^4\mbox{He}$ \cite{Helium}.

Conclusions concerning the shear viscosity per unit of entropy of the medium produced in Au+Au collisions at RHIC are based largely on comparisons of non-dissipative hydrodynamical calculations of the time evolution of collision systems with measurements of the particle production azimuthal anisotropy characterized by the elliptic flow coefficient $v_{2}$ \cite{Ref2,Elliptic}. These calculations describe the $v_{2}$ and momentum spectra measured in Au+Au collisions at $\sqrt{s_{NN}} = $~200 \mbox{GeV} well at midrapidity ($|\eta|<1.0$), low transverse momentum ($p_{T} < $ 1 \mbox{GeV}/{\it c}), and for mid-central collisions (impact parameter {\it b}~$\leq$~5~fm)\cite{Ref1,Elliptic,Ref5}.  A measure of fluidity is provided by the ratio of shear viscosity, $\eta$, to entropy density, $s$, henceforth referred to as $\eta/s$. It has been conjectured that the limit for all relativistic quantum field theories at finite temperature and zero chemical potential is close to the Kovtun-Son-Starinets (KSS) bound,   $\eta /s|_{KSS}  =  {(4\pi)}^{-1} \approx 0.08$ \cite{Ref2,Ref7}. Estimates of $\eta/s$ based on $v_{2}$, measured in Au+Au collisions at $\sqrt{s_{NN}}~=~$ 200 \mbox{GeV}, range significantly below the viscosity per unit of entropy ratio of superfluid ${}^4\mbox{He}$ and very close to the quantum limit \cite{Ref2,Elliptic,Gavin,Phenix}. Given the importance of viscosity in furthering our understanding of QCD matter, it is of interest to consider alternative measurement techniques to estimate the magnitude of $\eta/s$. Measurements of di-hadron correlations in heavy ion collisions, carried out as a function of the relative azimuthal particle emission angle, $\Delta\phi$, have  greatly advanced the studies of hot and strongly interacting matter at RHIC \cite{Trigger}. Indeed, studies of correlations between low and high $p_{T}$ particles have revealed the modification of away-side ($\Delta\phi \sim \pi$) jets and the formation of a longitudinally elongated near-side ($\Delta\phi \sim 0$) structure, known as the ridge, in central Au+Au collisions~\cite{STAR}. 
Meanwhile, low-$p_{T}$ di-hadron correlation studies reveal rich correlation structures, particularly on the away-side \cite{STAR}. However, the interpretation of these different measurements is nontrivial, and a number of competing models invoking different reaction mechanisms have been suggested to explain the data, each with relative success \cite{THEORY}. Thus, additional observables and measurements are required to discriminate fully among these competing models.

In this work, we present measurements of the differential extension of an integral observable {\it C} \cite{Gavin} in Au+Au collisions at $\sqrt{s_{NN}}$~=~200 \mbox{GeV}. The correlation function {\it C} is defined as follows:
\begin{linenomath*}
\begin{equation}
C(\Delta\eta,\Delta\phi) = \frac{{\left\langle {\sum\limits_{i = 1}^{n_1 } {\sum\limits_{i \ne j = 1}^{n_2 } {p_{T,i} p_{T,j} } } } \right\rangle }}{{\left\langle n \right\rangle _1 \left\langle n \right\rangle _2 }} - \left\langle {p_T } \right\rangle _1 \left\langle {p_T } \right\rangle _2 
\label{Eq:2}
\end{equation}
\end{linenomath*}
where $\left\langle {p_T } \right\rangle_k  \equiv \left\langle {\sum {p_{T,i } } } \right\rangle_k /\left\langle n \right\rangle_k$ is the average momentum,  the label {\it k} stands for particles from each event and the brackets represent event ensemble averages. $\left\langle {n } \right\rangle_k$ is the average number of particles emitted at $\left( {\eta _k ,\phi _k } \right)$. The indices {\it i} and {\it j} span all particles in a $\left( {\eta _k ,\phi _k } \right)$ bin. $\Delta\eta=\eta_1-\eta_2$ and $\Delta\phi=\phi_1-\phi_2$ are the relative pseudorapidity and azimuthal angle of measured particle pairs, respectively.

The correlation observable $C(\Delta\eta,\Delta\phi)$, defined above, is an extension of the number correlation function $R_2$  used in various studies~\cite{r2def}. By construction, it measures the degree of correlation between particles emitted at fixed relative pseudorapidity, $\Delta\eta$, and azimuthal angle difference, $\Delta\phi$, and is as such sensitive to various aspects of the collision dynamics. However, the explicit transverse-momentum weighing provides for additional sensitivity to discriminate and study soft (low $p_T$) vs. hard (high $p_T$) processes. Note that {\it C} differs structurally and quantitatively from the observables $\left\langle {\delta p_{T} \delta p_{T}} \right\rangle$ \cite{Gary} and $\Delta \sigma^{2}_{p_{T}}$ \cite{Estruct} previously reported by STAR. Differences  stem from the fact that {\it C} is sensitive not only to number density fluctuations, but also to $p_{T}$ fluctuations, and as such reflects the magnitude of in-medium momentum current correlations \cite{Gavin}. 

This study is based on an analysis of $8 \times 10^{6}$ minimum bias (MB) trigger events recorded by the STAR experiment in the year 2004 (RHIC Run IV). The MB trigger was defined by requiring a coincidence signal of two zero-degree calorimeters (ZDCs) located at $\pm$18 m from the center of the STAR Time Projection Chamber (TPC). Data were acquired with forward ($+z-$axis) and reverse ($-z-$axis) solenoidal magnetic field polarity with nominal field strength of $0.5$ \mbox{T}.  Collision centrality was estimated based on the uncorrected primary track multiplicity within $|\eta|<1.0$. Nine centrality classes corresponding to 0-5\% (most central), 5-10\% up to 70-80\% (most peripheral) of the total cross-section were used. A mean number of participants, $N_{part}$, is attributed to each fraction of the total cross-section using a Glauber Monte Carlo simulation \cite{Monte}. 

The analysis is restricted to charged-particle tracks measured in the TPC with $|\eta|<1.0$. Particles of interest for our measurement are those emerging from the bulk of the matter. Comparisons of RHIC data to hydrodynamic models show that the (near) equilibrium description only holds for particles with $p_{T} \leq 2$~\mbox{GeV}/{\it c}. For larger momenta, particle production is dominated by hard processes. Thus, we restrict this measurement to low $p_{T}$, i.e., with both particles in the range $0.2<p_{T}<2.0$~\mbox{GeV}/{\it c}. Tracks were selected on the basis of standard STAR quality cuts \cite{NetCharge}. To minimize acceptance effects, events were analyzed provided their collision vertex lay within a distance of $|z|<25$ cm from the center of the TPC. However, the particle acceptance exhibits a small dependence on the collision vertex position, which may introduce artificial correlations in the measurement of $C$. To avoid such effects, we measure $C$ independently for forward and reverse magnetic field settings in 20 vertex-{\it z} bins of width $\Delta z = 2.5$ cm in the range $-25<z<25$ cm. Then we average these measurements to obtain the correlation function. Track reconstruction inefficiencies for pairs with  $\Delta\eta \sim 0$, due to track crossing or merging in the TPC, are corrected for by performing a $p_{T}$ and charge sign ordered analysis of these pairs. Track pair losses occur when two tracks pass nearby one another and produce overlapping charge clusters in the TPC. For instance, with a forward magnetic field setting (i.e.  along the $+z$-axis), two positive charged particles, with   $p_{T,2}>p_{T,1}$, may cross in the TPC if emitted at pseudorapidity difference $\Delta\eta \sim 0$, and relative angle $\Delta\phi<0$ thereby resulting in pair losses for $\Delta\phi<0$. Pairs emitted with $\Delta\phi>0$ however tend to diverge in the TPC and thus are not subject to such losses. In symmetric A+A collisions, pair correlation functions are invariant under $\Delta\phi  \rightarrow - \Delta\phi$ reflection. The lost pair yield at $\Delta\phi<0$ may thus be corrected based on the yield at $-\Delta\phi$. Same-sign track pairs are recorded with $\Delta\phi = -|\Delta\phi|$ for $p_{T,1}>p_{T,2}$ and $\Delta\phi = +|\Delta\phi|$ otherwise. Pair yields measured for $-1.0<\Delta\phi<0$, are then substituted for those at $0<\Delta\phi<1.0$, thereby compensating for pair losses. A similar technique is used for unlike-sign pairs. However, no  corrections are made for track pairs with $|\Delta\eta|<0.032$ and $|\Delta\phi|<0.087$ radian (bin at the origin). These corrections change the amplitude of {\it C} by $< 1$\% in peripheral collisions and up to 4\% in central collisions.
The measurements of $C(\Delta\eta,\Delta\phi)$ reported in this work were constructed using 31 and 36 bins along the $\Delta\eta$ and $\Delta\phi$ axes respectively. We verified
that the results are  independent of the bin width.
\begin{figure}[!htb]
\centering
\resizebox{7.cm}{16.0cm}{\includegraphics{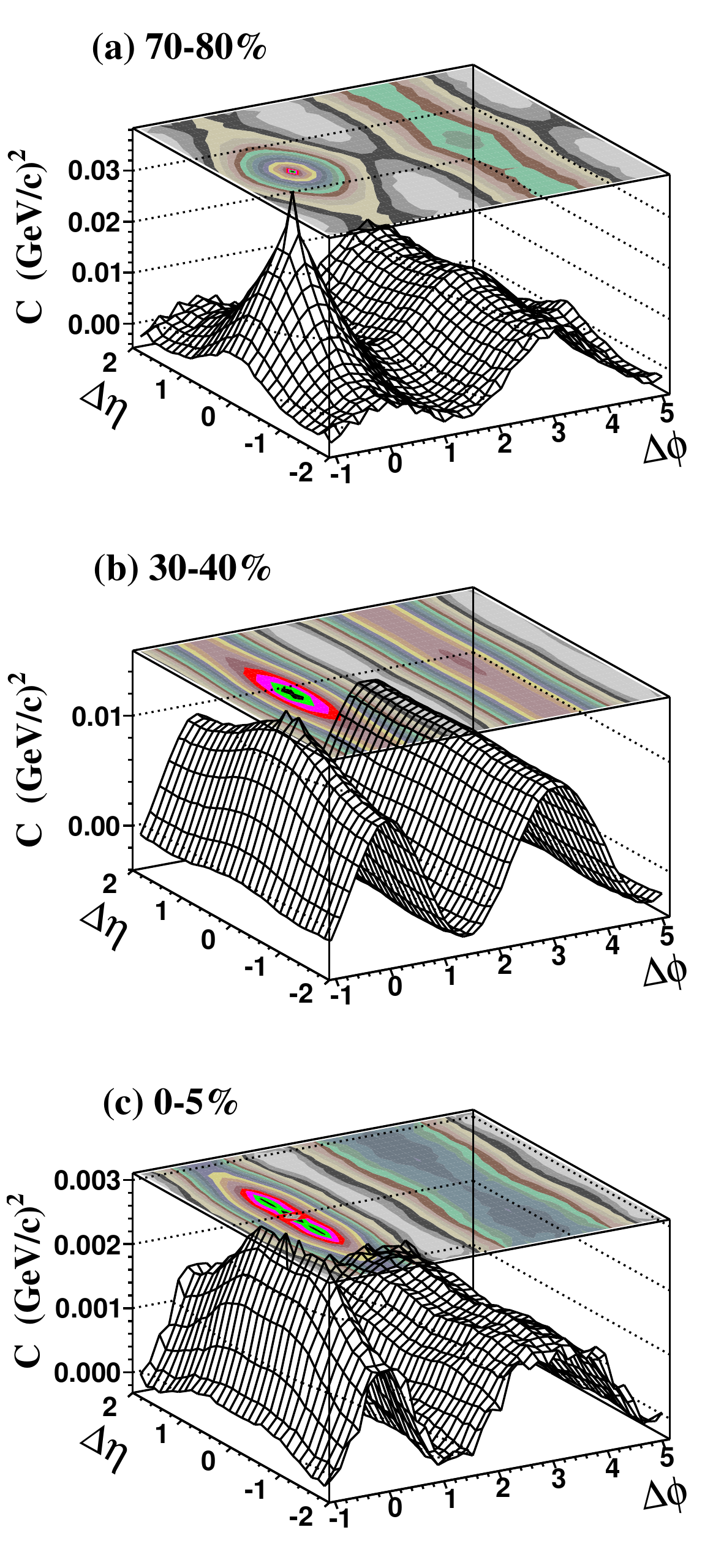}}
\caption[]{(Color online) Correlation function, ${\it C}$, shown for (a) 70-80\%, (b) 30-40\%, and (c) 0-5\% centrality in Au+Au collisions at $\sqrt{s_{NN}}$ = 200 \mbox{GeV}. {\it C} is plotted in units of $(\mbox{GeV}/{\it c})^{2}$, and the relative azimuthal angle $\Delta\phi$ in radians.}
\label{fig1}
\end{figure}
\begin{figure}[!htb]
\centering
\resizebox{7.0cm}{14.5cm}{\includegraphics{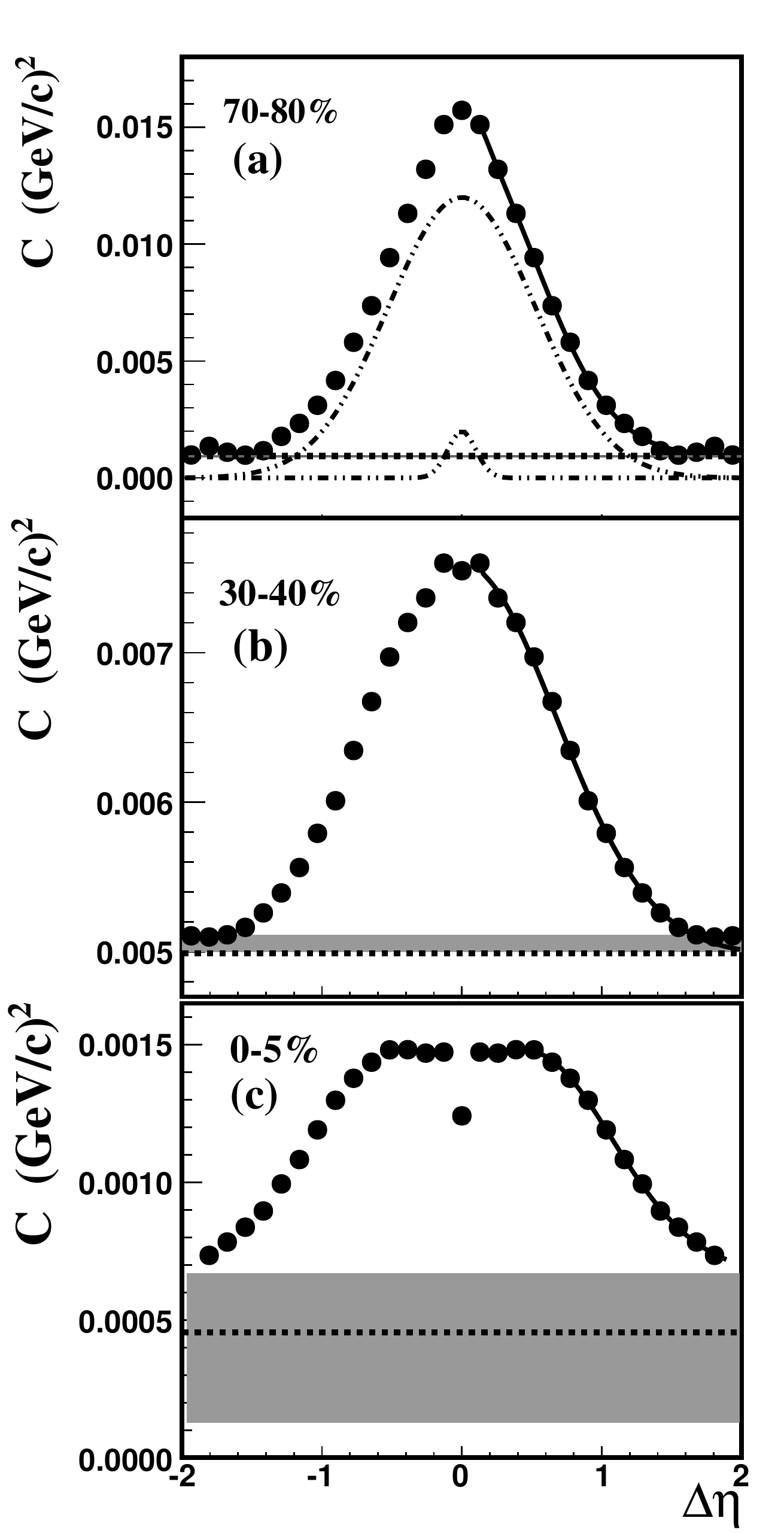}}
\caption[]{(a) Projection of the correlation function ${\it C}$, for $|\Delta\phi|<$ 1.0 radians on the $\Delta\eta$ axis for 70-80\% centrality, (b) 30-40\% centrality, and (c) 0-5\% centrality in Au+Au collisions at $\sqrt{s_{NN}}$ = 200 \mbox{GeV}. The correlation function {\it C} is plotted in units of $(\mbox{GeV}/{\it c})^{2}$. The solid line shows the fit obtained with Eq. \ref{Eq:3}. The dotted line corresponds to the baseline, {\it b}, obtained in the fit and shaded band shows uncertainty in determining {\it b}.}
\label{fig2}
\end{figure}

Figure \ref{fig1} presents the correlation function, {\it C}, for three representative collision centralities (a) 70-80\%, (b) 30-40\% and (c) 0-5\%.  Relative statistical errors range from 0.8\% in peripheral collisions to 0.9\% in the most central collisions at the peak of the distribution. Sources of systematic errors on the amplitude and shape of the correlation function include the collision centrality definition on the basis of primary particle multiplicity in the range $|\eta|<1.0$,  finite centrality bin width effects, loss of track reconstruction efficiency at $p_{T} < $ 0.5 \mbox{GeV}/{\it c}, B-field direction, and high TPC occupancy, as well as contamination of the correlation function from weakly decaying hadrons ($K_S^0$, $\Lambda$), conversion electrons, and HBT correlations.  A study of the effect of the centrality definition based on particle multiplicity in the range $|\eta|<0.5$,  $|\eta|<0.75$, and $|\eta|<1.0$ compared to that obtained with the ZDC energy reveals that the $|\eta|<1.0$ based centrality definition least biases the shape of {\it C} at large $\Delta\eta$. Uncertainties on the correlation yield associated with centrality boundaries and bin width vary from 10\% in peripheral to less than 1\% in the most central collisions. Contamination from weakly decaying particles and conversion electrons is estimated to contribute less than 2\% based on measured yields and known material budget of the detector.  HBT effects are essentially negligible, due to the large $p_{T}$ range used in the measurement. 

The overall strength of {\it C} decreases monotonically from peripheral to central collisions. In 70-80\% peripheral collisions, {\it C} exhibits a near-side peak centered at $\Delta\phi \sim \Delta\eta \sim $ 0 and a longitudinally extended away-side structure (i.e., broad in $\Delta\eta$) at $\Delta\phi \sim \pi$. This away-side structure largely results from effects associated with momentum conservation \cite{MomConservation}. In more central collisions, momentum conservation effects are diluted by increased particle multiplicities, and the near- and away-side observed correlation features may result from a superposition of several mechanisms possibly including resonance and cluster decays, radial flow effects, anisotropic flow effects, initial state fluctuations, and modified jet fragmentation. In mid-central collisions (30-40\%), the correlation function exhibits a sizable broadening of the near-side peak and the formation of a near-side ridge-like structure, as well as a strong elliptic flow, $\cos(2\Delta\phi)$, modulation \cite{V2}. In the most central collisions (0-5\%), we observe further longitudinal broadening of the near-side peak while the $\cos(2\Delta\phi)$ modulation and away-side structures have a much reduced amplitude. 
\begin{figure}[!htb]
\centering
\resizebox{8.0cm}{5.75cm}{\includegraphics[angle=90]{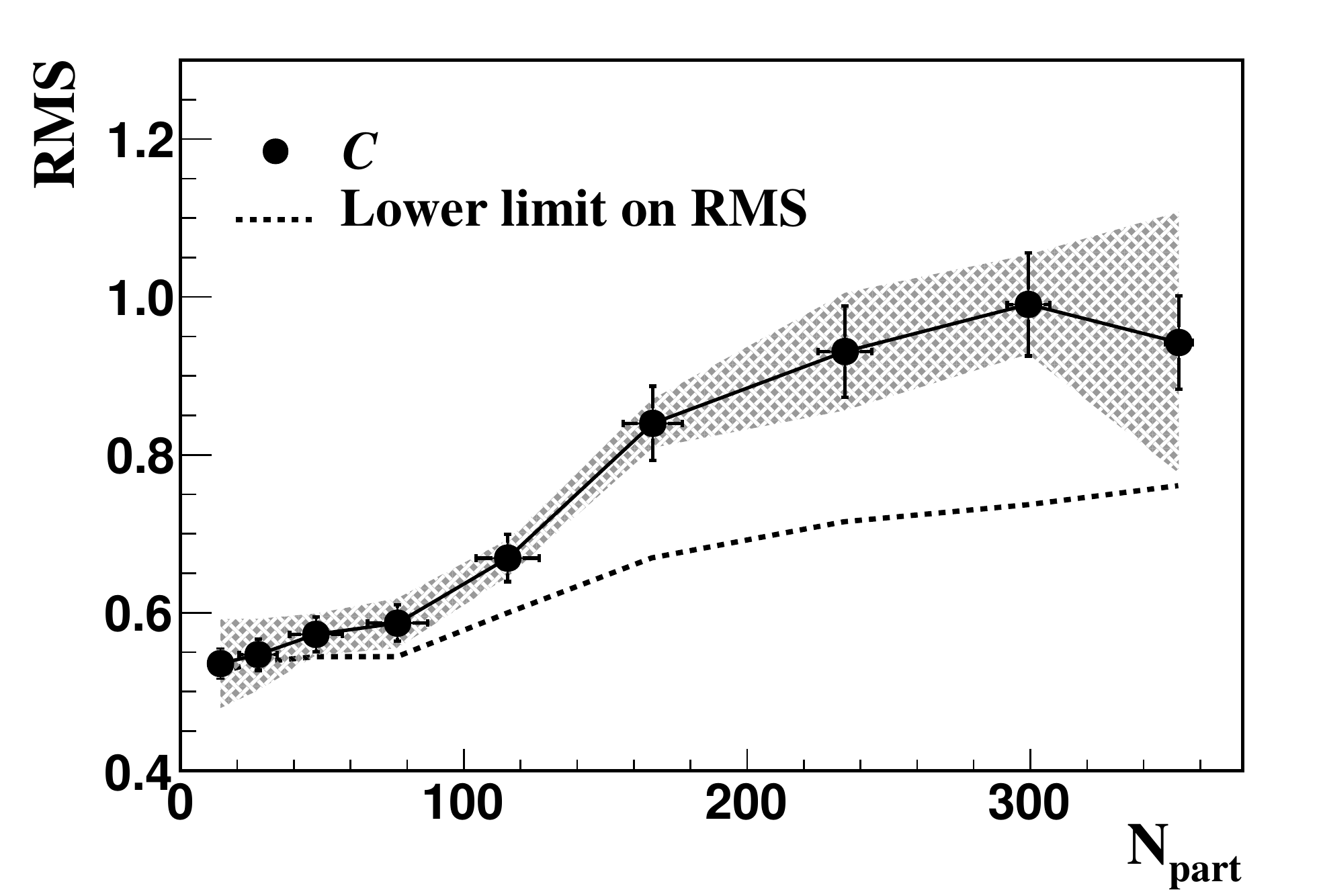}}
\caption[]{RMS as function of the number of participating nucleons for the correlation function {\it C}, for nine centrality classes in Au+Au collisions at $\sqrt{s_{NN}}$ = 200 \mbox{GeV}. The dotted line represents  a  lower limit estimate  of the RMS explained in the text and the shaded band represents systematic uncertainties on the RMS. }
\label{fig3}
\end{figure}

We next focus on the longitudinal broadening of $C$ with increasing $N_{part}$ based on $\Delta\eta$ projections in the range $|\Delta\phi|<$ 1.0 radians. Figures \ref{fig2}(a-c) show the projections for 70-80\%, 30-40\%, and 0-5\% centralities, respectively. The dip seen at $\Delta\eta \sim 0$ for 0-5\% central collisions (Fig. \ref{fig2}(c)) is a consequence of track merging occurring at $\Delta\phi \sim \Delta\eta \sim $ 0. We observe that the shape and particularly the width of the projections evolve with collision centrality. We characterize the widths of the distributions by calculating their RMS above a long range baseline, $b$,  assumed to be constant in the acceptance of our measurement. The baseline, $b$, is determined using the following ansatz to fit the projections:   
\begin{linenomath*}
\begin{equation}
\begin{array}{c}
 g\left( {b,a_w ,\sigma _w ,a_n ,\sigma _n } \right) = b + a_w \exp \left( { - \Delta \eta ^2 /2\sigma _w^2 } \right) \\ 
 + a_n \exp \left( { - \Delta \eta ^2 /2\sigma _n^2 } \right) \\ 
 \end{array}
\label{Eq:3}
\end{equation}
\end{linenomath*}
where $a_{w}$ and $a_{n}$ stand for the amplitude of wide and narrow Gaussians with widths $\sigma_{w}$ and $\sigma_{n}$, respectively.  The offset, narrow Gaussian,  wide Gaussian, and full fit are shown in Fig.~\ref{fig2}(a) for peripheral collisions. The fits have $\chi^2$ per degree of freedom values of order unity. The fits are used uniquely for the determination of the offset $b$. The amplitudes and widths of the Gaussians are not used in the remainder of this analysis. Uncertainties in the determination of the offset, $b$, are shown as dark gray shaded areas in Fig.~\ref{fig2}. 

Figure \ref{fig3} shows the RMS of the correlation function as a function of $N_{part}$. 
Vertical lines indicate statistical errors whereas systematic uncertainties on the RMS are indicated by the gray shaded  band. Systematic uncertainties arise from several sources. The correlation width exhibits small instrumental dependencies on the magnetic field direction, and the collision vertex position of the order of 3\% and 4\% respectively in most central collision and much smaller in peripheral collisions. Track merging corrections, discussed above, account for particles losses at  $|\Delta\eta|\sim 0$, $|\Delta\phi| < 1.0 $ and lead to negligible, $\ll 1$\%, systematic errors on the RMS of the distributions. The correction technique used does not account for losses at $|\Delta\eta|<0.032$ and $|\Delta\phi|<0.087$ radian (bin at the origin) which are most severe in 0-5\% central collisions. This bin
is also subject to contamination from $e^+e^-$ pairs resulting from photon conversions within the apparatus. We estimated the latter two effects introduce small systematic uncertainties, $< 2$\%, on the RMS of the correlation functions. The largest source of systematic uncertainties 
stems from the baseline determination and the lack of knowledge of the correlationÕs long $\Delta\eta$ range behavior, particularly in central collisions. In order to study these effects, we first estimated a lower bound of  RMS values, shown as a dotted line in Fig.~3, by setting the offset equal to the value of the correlation signal at $\Delta\eta = 2.0$. This simplistic calculation shows that the RMS exhibits a monotonic growth from peripheral to central collisions. In peripheral collisions, the correlation peak stands atop an approximately flat background but in most central collisions the peak is manifestly broader than the acceptance and this simple estimate is therefore incorrect. We thus used  Eq. \ref{Eq:3} and systematically studied fits  for various number of  parameters and fit ranges. Estimated systematic uncertainties on the offset are shown as gray bands in Fig.~2. Uncertainties on the offset and  shape of the distribution, particularly in central collisions, lead to systematic uncertainties on the RMS ranging from 10 \%  in peripheral collisions to 15 \%  in most central collisions. The above  systematic uncertainties are summed in quadrature and shown as a gray shaded band in Fig.~3. The RMS exhibits a modest increase in the range $N_{part} < 100$ which may in part result from long range multiplicity fluctuations and from incomplete system thermalization achieved in small collision systems. The RMS rises rapidly in the range $100<N_{part} < 250$ after which it levels off. 

According to \cite{Gavin}, shear viscosity should dominate the broadening of the correlation function for sufficiently large and nearly thermalized collision systems. It should thus be possible to utilize the observed broadening to estimate the viscosity of the matter produced in these collisions. However, jets and jet quenching could also in principle contribute to changes in the shape and broadening of the width of the correlation function with varying collision centralities. To examine this possibility, we repeated our analysis in the $0.2<p_{T}<1.0$~\mbox{GeV}/{\it c} and $0.2<p_{T}<20.0$~\mbox{GeV}/{\it c} ranges. Our study shows that particles accepted between $0.2<p_{T}<20.0$~\mbox{GeV}/{\it c} produce essentially identical widths in peripheral collisions. In central collisions, RMS reduces by $\sim$7\% from the RMS widths obtained for the $p_{T}$ selection $0.2<p_{T}<2.0$~\mbox{GeV}/{\it c}. However, lowering the upper $p_{T}$ cut to 1.0 \mbox{GeV}/{\it c} ($0.2<p_{T}<1.0$~\mbox{GeV}/{\it c}) does not change the widths within statistical errors for $0.2<p_{T}<2.0$~\mbox{GeV}/{\it c} range for the most central collisions, and decreases the widths by $\sim$10\% in peripheral collisions. We conclude that broadening effects associated with jets or jet quenching are thus likely limited to less than a 10\% effect on the RMS from peripheral to central collisions. We thus proceed to 
estimate the shear viscosity per entropy of the matter produced in central collisions based on the following formula from Ref. \cite{Gavin}.
\begin{linenomath*}
\begin{equation}
\sigma _c^2  - \sigma _0^2  = 4\frac{\eta}{T_{c}s} \left( {\tau _{0}^{ - 1}  - \tau _{c,f}^{ - 1} } \right)
\label{Eq:1}
\end{equation}
\end{linenomath*}
where $\sigma_c$ and $\sigma_0$ stand for the longitudinal widths of the correlation function in central collisions and at formation time, respectively. $\tau_{0}$ refers to the formation time and $\tau_{c,f}$ is  the kinetic freeze-out time at which particles have no further interactions \cite{TIME}. {\it $T_{c}$} stands for a characteristic temperature of the system through its evolution, and is here taken to be the critical temperature. We proceed by assuming that viscous broadening dominates the increase in {\it C} with increasing centrality observed in this analysis and utilize Eq. \ref{Eq:1}  to estimate $\eta/s$. We estimate $\sigma_{o}$ = 0.54 $\pm$ 0.02({\it stat.}) $\pm$ 0.06({\it sys.}) by extrapolating the RMS width of {\it C} to $N_{part} \sim 2$. The RMS value for most central collisions is $\sigma_{c} = 0.94 \pm 0.06(stat.) \pm 0.17(sys.)$. Using commonly accepted estimates of 1 fm/$c$, 20 fm/$c$, and 170 MeV~\cite{Gavin} for the formation time, central collision freeze-out, and effective temperature, we obtain a value of $\eta/s=0.13 \pm 0.03$.  Inclusion of systematic uncertainties on the widths leads to a range of $\eta/s~=~0.06~-~0.21$. Figure \ref{fig4} shows $\eta/s$ as a function of $\tau_{0}^{-1} - \tau_{c,f}^{-1}$ and provides an estimate of theoretical uncertainties based on a  literature survey of theoretical estimates for  $\tau_{0}$ and $T_{c}$.   $\tau_{0}$ is typically assumed to be in the range 0.6 - 1.0 fm/{\it c} (e.g., \cite{Gavin,TIME,Song}). Here, we have assumed that the broadening of {\it C} is entirely due to viscous effects. Given that other (unknown) dynamical effects could perhaps also lead to the correlation function broadening, we conclude that our measurement provides an upper limit. 
Based on the statistical and systematic uncertainties of our measurement (using one standard deviation) and caveats of the used theoretical model, and using the ranges $150 < T_c< 190$  MeV and
$ 0.6< \tau_{0}^{-1} - \tau_{c,f}^{-1} <$ 1.6 (fm/c)$^{-1}$, we derive an upper limit of order $\eta/s \sim 0.3$ .
 
\begin{figure}[!htb]
\centering
\resizebox{8.0cm}{5.75cm}{\includegraphics[angle=90]{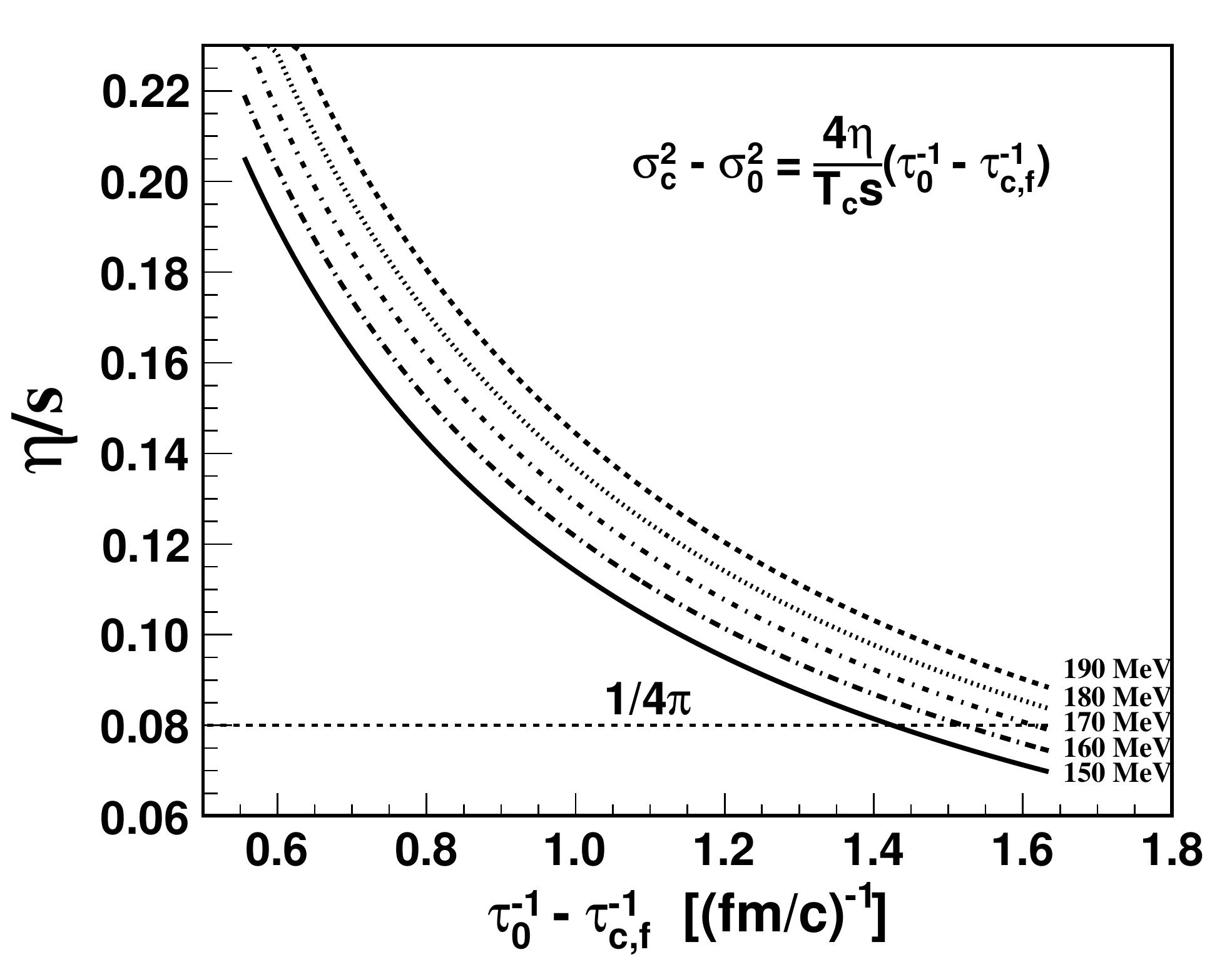}}
\caption[]{$\eta/s$ as a function of $\tau_{0}^{-1} - \tau_{c,f}^{-1}$ and $T_c$. $\tau_{0}$ and $T_c$ vary from $0.5<\tau_{0}<1.5$ fm/$c$ and $150<T_c<190$ MeV, respectively.}
\label{fig4}
\end{figure}

In summary, we presented first measurements of the differential transverse momentum correlation function {\it C} from Au+Au collisions at $\sqrt{s_{NN}}$ = 200 \mbox{GeV}.   In peripheral collisions, $C$  has a shape qualitatively similar to that observed in measurements of number density correlations, with a relatively narrow near-side peak near $\Delta\eta \approx \Delta\varphi \approx 0$ and a  longitudinally broad away-side \cite{Trigger,Estruct}.  We find that the near-side peak  progressively broadens with increasing number of collision participants while the overall strength of the correlation function decreases monotonically.  These results may be used to further constrain particle production and correlation models. We used the observed longitudinal broadening to estimate $\eta/s$ of the matter formed in central Au+Au collisions.  Considering systematic uncertainties in the determination of correlation widths, particularly in central collisions, and assuming somewhat conservative estimates of the temperature, formation and freeze-out times, we obtain a range of $\eta/s$ = 0.06 - 0.21. This result is remarkably close to the KSS bound, ${(4\pi)}^{-1}$, and is consistent with results obtained from hydrodynamical model comparisons to elliptic flow data \cite{Elliptic}. 

We thank the RHIC Operations Group and RCF at BNL, the NERSC Center at LBNL, and the Open Science Grid consortium for providing resources and support. This work was supported in part by the Offices of NP and HEP within the U.S. DOE Office of Science, the U.S. NSF, the Sloan Foundation, the DFG cluster of excellence `Origin and Structure of the Universe' of Germany, CNRS/IN2P3, STFC and EPSRC of the United Kingdom, FAPESP CNPq of Brazil, Ministry of Ed. and Sci. of the Russian Federation, NNSFC, CAS, MoST, and MoE of China, GA and MSMT of the Czech Republic, FOM and NWO of the Netherlands, DAE, DST, and CSIR of India, Polish Ministry of Sci. and Higher Ed., Korea Research Foundation, Ministry of Sci., Ed. and Sports of the Rep. Of Croatia, Russian Ministry of Sci. and Tech, and RosAtom of Russia.

\end{linenumbers}
\end{setpagewiselinenumbers}
\end{document}